\documentclass[pre,aps,amsmath,amssymb,superscriptaddress,longbibliography,notitlepage,twocolumn]{revtex4-1}
\usepackage{graphicx,multirow,outlines,ulem}
\usepackage{color}
\usepackage{hhline}
\usepackage{physics}
\usepackage{hyperref}
\usepackage{bm}

\usepackage[mathscr]{eucal}

\usepackage{tikz}
\usepackage{xcolor}

\newcommand{\bo}{\begin{outline}}
\newcommand{\eo}{\end{outline}}

\newcommand{\qed}{\nobreak \ifvmode \relax \else
      \ifdim\lastskip<1.5em \hskip-\lastskip
      \hskip1.5em plus0em minus0.5em \fi \nobreak
      \vrule height0.75em width0.5em depth0.25em\fi}

\begin{document}

\title{{\bf PT}-Symmetry Breaking Transitions in  Polymeric Systems }
\author{Tanmoy Pal}
\affiliation{Department of Biochemistry, University of Toronto, Toronto, Ontario M5S 1A8, Canada}
\author{Ranjan Modak}
\affiliation{ Department of Physics, Indian Institute of Technology Tirupati, Tirupati, India~517619} 
\author{Bhabani Prasad Mandal}
\affiliation{Department of Physics, Institute of Science, BHU, Varanasi 221005, India}
\begin{abstract}
We show that classical DNA unzipping transition which is equivalently described by quantum mechanical localization-delocalization transition in the ground state of non-Hermitian single impurity Hatano-Nelson Hamiltonian is underpinned by 
generalized parity ({\bf P})-time reversal ({\bf T}) symmetry breaking transition. 
We also study the one-dimensional discretized version of Hatano-Nelson model in the presence of the single impurity and random disorder on a finite-size lattice. 
These discrete models are useful to study unzipping of a single adsorbed polymer from a surface.
Our results show that the discrete models also undergo a phase transition from a PT unbroken phase to a broken phase. Interestingly, the generalized PT phase transition points coincide with the localization-delocalization transition for  continuum as well as lattice models.

\end{abstract}
\maketitle
\section{Introduction}
Hermiticity requirement of quantum mechanical Hamiltonians 
can be relaxed  with a physical but less constraining condition of  combined  {\bf P} and {\bf T} invariance 
~\cite{Bender_1998, Bender_2002}. 
Such non-Hermitian {\bf PT} symmetric systems 
generally exhibit a {\bf PT} breaking phase transition that separates two regions of a Hermiticity breaking parameter: 
(i) region of the unbroken {\bf PT} symmetry in which the entire energy spectrum is real 
and corresponding eigenstates respect {\bf PT} symmetry and (ii) a region of the broken {\bf PT} symmetry in 
which the whole spectrum (or a part of it) appears as complex conjugate pairs and the eigenstates of the systems 
do not respect {\bf PT }symmetry~\cite{Bender_2002,Bender_2007,Mostafazadeh_2007,Ju_2019,Bender_1998,shukla2022heisenberg, Modak_2021, MANDAL2015185, KHARE200053, MANDAL20131043, RAVAL2019114699}.  {\bf PT}-symmetric quantum 
theories are applicable in various branches of physics including open quantum systems, scattering theory, optics \cite{musslimani_2008,guo_2010,west_2010} etc.  The consequence of {\bf PT} phase transition has been 
experimentally observed in non-linear optics where {\bf PT} symmetric complex potentials can be realised 
through complex refractive index and several physical processes 
are known to follow Schrödinger like equations~\cite{Ozdemir_2019}. This created an avalanche in the 
study of non-Hermitian quantum theories \cite{Bender_2007,MOSTAFAZADEH_2010}. To buttress the general 
applicability of {\bf PT}-symmetric quantum theories, observation of {\bf PT} phase transition in physical 
systems other than optical media is thus very useful.

Before the advent of {\bf PT}-transition, Hatano and Nelson (HN) introduced a class of non-Hermitian but real 
model Hamiltonians through which they described pinning-depinning transition of a flux-line to a linear defect 
in Type-II superconductors \cite{HatanoNelson1996PRL, HatanoNelson1997PRB}.  Whether such HN-models could 
show {\bf PT} transitions is a natural question to ask. A hallmark of the HN-Hamiltonians 
is the presence of imaginary vector potentials combined with spatially short-range on-site potential. HN-Hamiltonians proved to be applicable to a variety of physical systems. In a rather surprising development, 
DNA unzipping was shown to be a classical equivalent of the localization-delocalization transition in the ground state 
of a HN-Hamiltonian that did not contain any space dependent random energy contribution \cite{LubenskyNelson2000,SMB2000}. DNA unzipping is the first step towards 
commencement of several life-sustaining biophysical processes including DNA-replication,  RNA transcription etc. \cite{SINDEN19981, Watson2007}. 
{\it In vivo}, DNA unzipping is carried out by a special class of enzymes 
called ``helicase'' by means of applying mechanical force that is experimentally measured to be $\sim$ 10-15 pN 
\cite{BOCKELMANN20021537,Bocklemann1997PRL, UmateHelicase}. {\bf PT} phase transition in a modified interacting HN-Hamiltonian has 
been studied recently~\cite{ZhangPhysRevB2022}. However, whether the HN-Hamiltonians that are relevant to DNA unzipping support {\bf PT} phase transition is still unknown. It is well known that unzipping 
of a double stranded DNA (dsDNA) is a genuine first order phase transition in which dsDNA unwinds completely 
when the applied unzipping force exceeds a critical value~\cite{KapriSMB2004, Garima2011JCP, KUMAR20101, TP_SMB_PRE_2016,PalPRE2022,DebjyotiSMBPRE}.  Nevertheless, a crucial question still remains unanswered: what is the 
accompanied symmetry breaking in unzipping transition and what is its quantum picture equivalent? A study of 
the HN type Hamiltonians in the framework of  {\bf PT} symmetric  non-Hermitian quantum mechanical formulation thus presents a unique opportunity 
to answer these questions in a single set up.
Moreover, HN-models are also relevant to the problem of mechanical force induced desorption of an adsorbed polymer from a surface ~\cite{orlandini2004adsorption,mishra2004force}, which has many applications such as lubrication, adhesion, surface protection, coating of surfaces, and wetting.  For reasons that we make apparent in the Section-III of this work, we are specifically interested to study the problem of unzipping an adsorbed polymer in a random/dirty medium. Because of the presence of randomness,  we use a lattice version of an one-dimensional HN Hamiltonian~\cite{KapriSMBUnzipDirtyEnv} to study such systems for convenience. Drawing motivations from the DNA unzipping problem, we also analyse these discrete models in the {\bf PT}-symmetric quantum mechanics picture.  

In this paper, we study two HN systems described by Dirac delta and random 
potentials. The delta potential can be utilized to model DNA unzipping and flux-depinning in Type-II superconductors, 
and the random potential can mimic the disordered medium which  potentially can describe a physically realizable situation for polymer adsorption or for flux
lines with columnar pins. Using analytical 
and numerical methods, we show that these models support {\bf PT} symmetry 
breaking phase transitions.  Importantly, we thus realize that DNA unzipping or even unzipping of adsorbed polymers is a spontaneous {\bf PT} symmetry 
breaking transition. {\it In vitro}, DNA unzipping has been realized at room temperature~\cite{BOCKELMANN20021537,Bocklemann1997PRL, ClaudiaDanilowicz1, ClaudiaDanilowicz2, EssevazRoulet, Anselmetti}.This means that dsDNA can be a potential testing ground for {\bf PT} symmetric quantum mechanics at room temperature beside low temperature quantum 
systems.

\section{Continuum Model}
 Here, we revisit the existing works on HN Hamiltonian in DNA context and simply restate the essential results relevant to our discussion \cite{HatanoNelson1996PRL, HatanoNelson1997PRB, SMB2000, LubenskyNelson2000}. For simplicity, let's model a dsDNA as two Gaussian polymers of contour length $N$ interacting through a short-range potential that acts only at the same contour length, thus mimicking native base pairing. Moreover, one end of the dsDNA is anchored at the origin and a time-independent constant force $\bm{g}$ is applied at the free ends of the two polymers in opposite direction. Up to an additive center of mass term the resulting Hamiltonian can be written as 
\begin{equation}
\label{eq02}
\frac{H}{k_B T}=\int^{N}_{0}d\tau\left[\frac{\epsilon}{2}\left(\frac{\partial\textbf{r}(\tau)}{d\tau}\right)^2 
            + V(\textbf{r}(\tau)) - \bm{g}.\frac{\partial\textbf{r}(\tau)}{\partial\tau}\right].
\end{equation}
Eq.~\eqref{eq02} describes a single polymer of elastic constant $\epsilon$ under a force $\bm{g}$ 
in a \textit{short range} potential $V({\bf r})$ where $\tau$ is the contour length variable and $\textbf{r}(\tau)$ is the $\tau$ 
dependent segment coordinate. The partition function $Z=\int\mathcal{DR}\exp(-H/k_B T)$ is obtained by 
summing over all polymer configurations~\cite{Comment1}. 
By treating $\tau$ as a time-like coordinate, the partition function $Z$ can be interpreted 
as the transition amplitude of a fictitious quantum particle through the imaginary time mapping 
$\tau\rightarrow \mathrm{i} t$ and $k_B T$ identified as $\hbar$. The Hamiltonian of the corresponding quantum 
particle is given by 
\begin{equation}
\label{eq03}
H_{\rm q}(\bm{g}) = \frac{1}{2\epsilon}({\bf p} + \mathrm{i}\bm{g})^2 + V({\bf r}),
\end{equation}
where $\epsilon$ is the mass of the quantum particle and ${\bf p}$ is canonical momentum. Thee unzipping force 
${\bf g}$ appears in $H_q({\bf g})$ as an imaginary vector potential, introducing non-Hermiticity.  Without loss of 
generality, we set $\hbar = \epsilon = 1$. Note that 
similar classical to quantum mapping have been used to study Efimov effect in alternative DNA 
conformations~\cite{efi1,efi2,efi3,efi4}.
%
\begin{figure}[t]
  \begin{center}
   \includegraphics[width=0.5\textwidth]{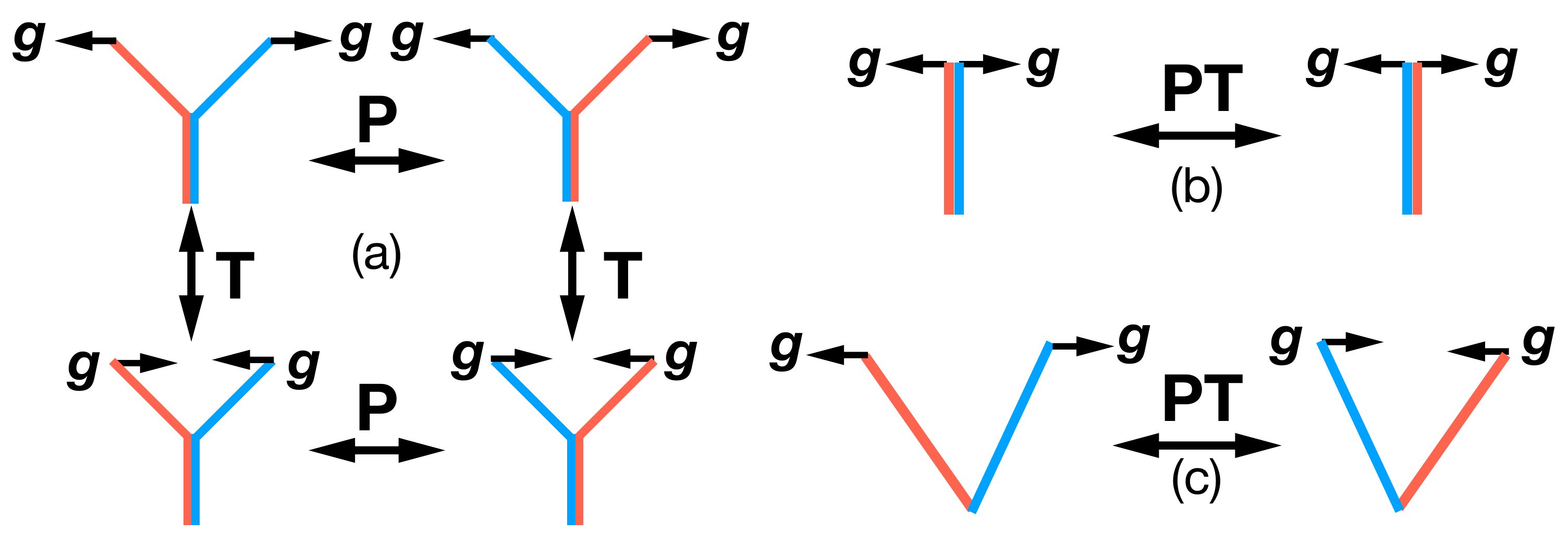}
   \caption{Schematic representation of {\bf PT}-transformation applied to a dsDNA. Two single strands are 
   identified by the colours red and blue, and the arrows show the direction of the applied unzipping force of 
   magnitude $\bm{g}$. (a) Changes for partially opened dsDNA under {\bf P} and {\bf T} transformations. (b) Zipped dsDNA 
   state remains invariant under {\bf PT} transformation.(c) Unzipped dsDNA state changes under {\bf PT} 
   transformation. }\label{fig01}
 \end{center}
\end{figure}

We define {\bf P} and {\bf T} transformations in the context of dsDNA. {\bf P} transformation 
exchanges the position coordinates of the native base pairs keeping the relative force 
between the strands unchanged whereas 
 {\bf T} 
 reverses the direction 
of the applied unzipping force keeping the relative distance between base pairs unchanged.  The effects of these {\bf P} and {\bf T} 
transformations  on zipped, partially zipped and unzipped dsDNA are schematically demonstrated in Fig.~\ref{fig01}.  We can readily see 
from Fig.~\ref{fig01}b and \ref{fig01}c that the totally zipped state of dsDNA remains invariant under {\bf PT} but 
partially or completely unzipped state of dsDNA changes under {\bf PT}. This indicates that dsDNA unzipping 
could be seen as a classical {\bf PT}-breaking transition. 

 In the quantum domain   {\bf P} 
and {\bf T} transformations are defined as
\begin{eqnarray}
\label{eq04}
{\bf P}&:& {\bf p} \rightarrow - {\bf p},~\ \ {\bf r} \rightarrow -{\bf r},~ \ \  \bm{g}: 
\rightarrow \bm{g},\quad \text{and,}\nonumber\\ 
{\bf T}&:& {\bf p} \rightarrow  -{\bf p},~ \ \  \bm{g} \rightarrow -\bm{g},~ \ \  {\bf r}: 
\rightarrow {\bf r},~\ \ \mathrm{i} \rightarrow -\mathrm{i}.
\end{eqnarray}
Note that both the Hamiltonians Eqs.~\eqref{eq02} and \eqref{eq03} remain invariant under {\bf PT}.

We restrict ourselves to one dimension to further simplify the problem at hand. Exact solutions for $V(x)=-V_0\delta(x)$, where $V_0>0$, have been obtained by Hatano \textit{et. al.}~\cite{HatanoNelson1997PRB} with periodic boundary condition. For $g<g_{\rm c}=V_0$, the ground state remains localized with $g$ independent real energy eigenvalue 
$E_{\rm gs}=-V^2_0/2$. The ground state delocalizes for $g>g_{\rm c}$. In the classical picture, the localized/delocalized 
ground state corresponds to the zipped/unzipped dsDNA~\cite{SMB2000, LubenskyNelson2000}. The excited-state energies are given by, $E_{\rm ex}=(k_1+\mathrm{i}k_2)^2/2$ where up to $O(L^{-1})$, 
\begin{eqnarray}
(k_1, k_2) &=&\left(\frac{2n\pi}{L}, g+\frac{1}{L}\ln\frac{g}{g-g_c}\right), \forall g>g_{\rm c},\; \text{and},\nonumber\\
(k_1, k_2) &=&\left(\frac{(2n+1)\pi}{L}, g+\frac{1}{L}\ln\frac{g}{g_c-g}\right), \forall g<g_{\rm c},\nonumber
\end{eqnarray}
where $n$ are integers and $L$ is system size. $n=0$ correspond to the 1st excited state and delocalized ground state for $g<g_c$ and $g>g_{\rm c}$, respectively. The ground state wavefunction for $g<g_{\rm c}$ reads as, 
\begin{eqnarray}
 \psi^{\rm R}_{\rm gs}(x,g<g_{\rm c})&\propto& \exp[-(V_0-g)x] ~~~~x>0, \nonumber \\
 &\propto&\exp[(V_0+g)x]~~~~~~~x<0. \nonumber
\end{eqnarray}
The ground state wavefunction for $g>g_{\rm c}$ is, 
\begin{eqnarray}
 \psi^{\rm R}_{\rm gs}(x,g>g_{\rm c})&\propto& e^{ik_1x} ~~~~~~~~~~~~~~~~~~~~~~x>0, \nonumber \\
 &\propto&e^{ik_1x}+Ae^{-ik_1x+2gx}~~~~x<0., \nonumber
\end{eqnarray}
where $A$ is a constant. It is easy to check that under the {\bf PT}-transformation rule  Eqn.~\eqref{eq04},
$\psi^{R}_{\rm gs}(x,g<g_{\rm c})$ remains invariant, while $\psi^{\rm R}_{\rm gs}(x,g>g_{\rm c})$ does not. 
In thermodynamic limit $L \to \infty$,  while the ground state energy is real for all $g$, only $\psi^{\rm R}_{\rm gs}(x,g<g_{\rm c})$ remains eigenstate of the {\bf PT} operator. This implies that the ground state of the continuum Hatano-Nelson model goes through a {\bf PT} transition. Hence, the localization-delocalization transition and DNA unzipping transition, 
both corresponds to the {\bf PT}-symmetry breaking transition. 
It is to be noted that $n$ can be as large as $L-1$. Thus, in the thermodynamic limit, $E_{\rm ex}$ remains complex for all $g$ which indicates that the {\bf PT}-transition for 
the studied model is restricted to the ground state only.

\section{Discrete Model} 
Having shown that the ground state of the 1D Hatano-Nelson model in continuum displays generalized PT
transition, here we study the discretized version of that
model on a finite lattice. Given in previous studies, the role of disorder has been
investigated, especially in Ref.~\cite{LubenskyNelsonPRE} the potential was varied randomly as a function of the time-like
variable to mimic a random DNA sequence, and in Ref.~\cite{KapriSMBUnzipDirtyEnv}, randomness was introduced in the space-like direction to study the
adsorption of a single-stranded polymer to a surface, we also introduce an on-site disorder in the discretized version of the Hatano-Nelson model. The Hamiltonian reads as, 
\begin{eqnarray}
\mathcal{H}_{\rm q} &=&\sum_{j}(e^{g}{c}^{\dag}_j{c}_{j+1}+e^{-g}{c}^{\dag}_{j+1}{c}_{j})-\mu\sum_{j} V_j{n}_j-V_0{n}_{L/2},\nonumber \\
\label{nonint_model}
\end{eqnarray}
numerically on a one-dimensional lattice of size $L$, where
${c}^{\dag}_j$  (${c}_j$) is the fermionic creation (annihilation) operator at site $j$, ${n}_j ={c}^{\dag}_j{c}_{j}$ is the number operator, on site potential $V_j$ is real and positive and lattice spacing set to unity. 
We investigate two models in details, Model A: $\mu=0$
(discrete version of the single-impurity HN model), and Model B: 
$V_j$ is sampled from uniform random distribution between an interval $[-V_0,V_0]$, 
a model that can potentially describe a physically realizable situation for polymer adsorption or for flux
lines with columnar pins. Note that we have also checked the validity of our results for Gaussian random disorder. 
Unlike Model A, on-site potential for Model B is not parity symmetric. However, for all our calculations  
we do an averaging over many disorder realizations and the ensemble $\{\mathcal{H}_{\rm q}\}$ of all
disorder realizations of the model B is {\bf PT} symmetric~\cite{PhysRevB.106.L060301}.In the Hermitian limit, the model B 
in one-dimension displays Anderson localization for any infinitesimal amount of disorder~\cite{anderson.1958}. 
Periodic boundary conditions and fixed $V_0=20$ have been used for all of our calculations.

\begin{figure}
    \centering
    \includegraphics[width=0.48\textwidth]{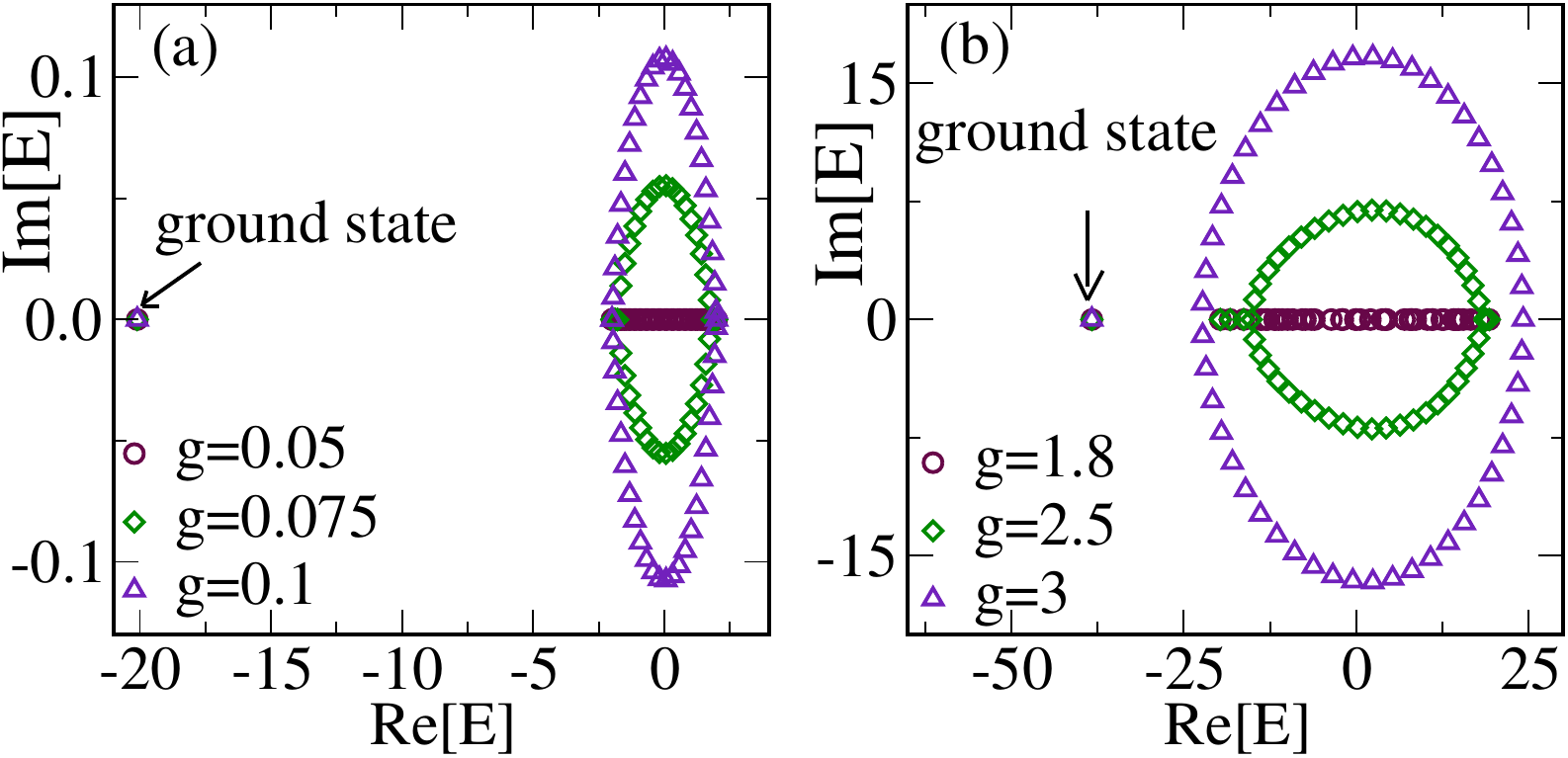}
    \caption{Variation of the imaginary part of energy eigenvalues with the real 
                    part of energy eigenvalues for (a) model A at $g=0.05$, $g=0.075$, and $g=0.1$. 
                   (b) model B for $g=1.8$, $g=2.5$, and $g=3$ 
                   All results are for $L=50$.}
    \label{fig:energy}
\end{figure}

First, we show the energy spectrum of model A and model B for $L=50$ in Fig.~\ref{fig:energy} in the complex plane.  We plot the real part of the energy eigenvalues in $x$ axes along with its corresponding imaginary counterpart in the $y$ axes. We find in Fig.~\ref{fig:energy} that the energy eigenvalues are completely real for $g=0.05$ (for model A) and $g=1.8$ (for model B). On the other hand, they become complex for $g=0.075, 0.1$ for model A and for $g=2.5, 3$ for model B. Real to complex transitions in eigenvalues provide hints that maybe both models go through {\bf PT} transitions. On top of that, the imaginary part of the eigenvalues remains symmetric about $x$ axes, and that proves the complex eigenvalues appear as a complex conjugate pair in the spectrum, which is also a property of the  {\bf PT} broken phase. Finally, we would also like to point out a striking feature, i.e., the ground state energy remains real for all values $g$ that have been studied here.

\begin{figure}
    \centering
    \includegraphics[width=0.48\textwidth]{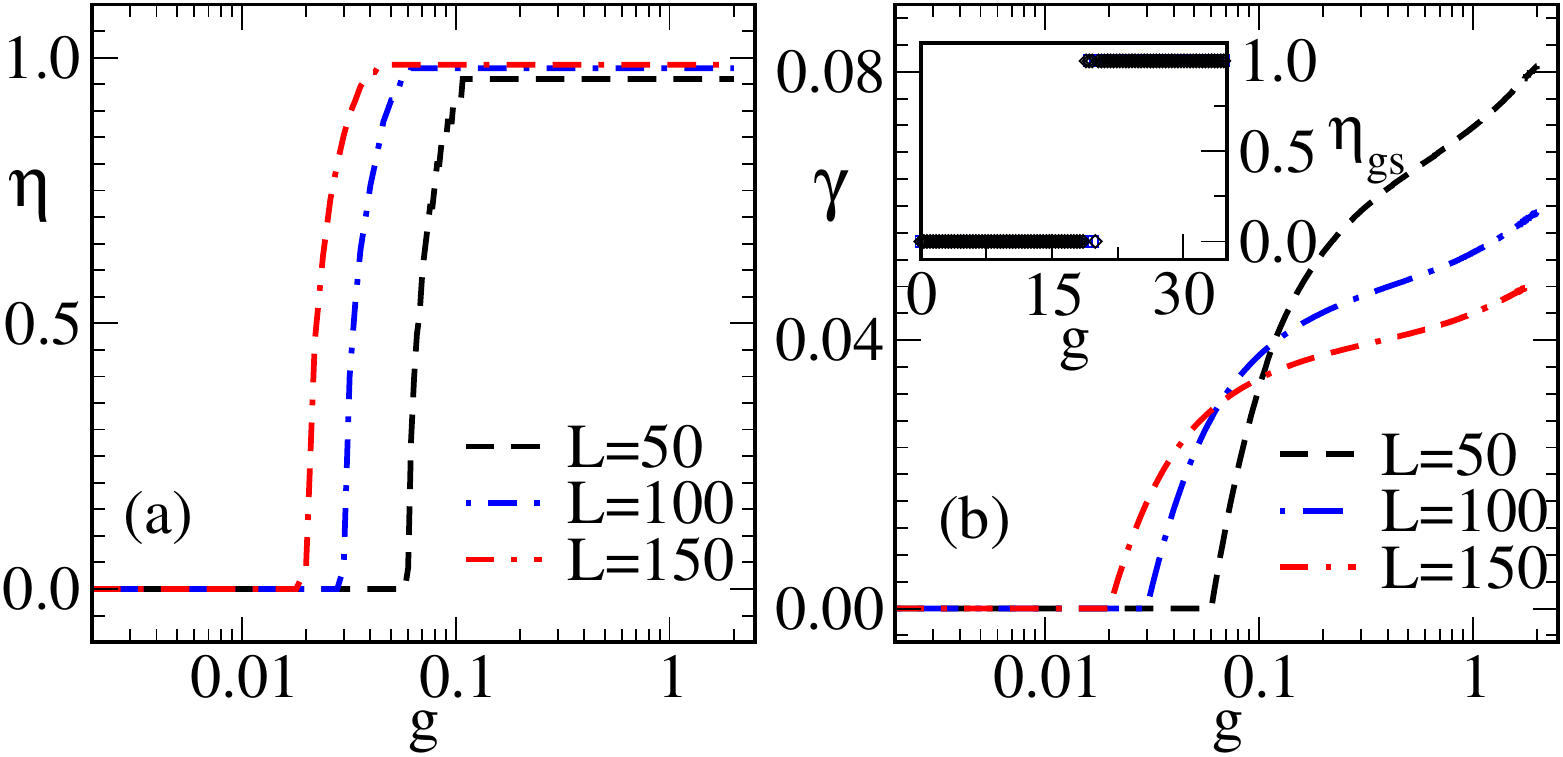}
    \caption{(a) shows the variation of $\eta$ with $g$. (b) shows the variation of $\gamma$ with $g$ for different values of $L=50$, 100, 150, and for Model A. Inset of right panel shows variation of $\eta_{gs}$ for the ground state as a function of $g$.}
    \label{fig1}
\end{figure}

\begin{figure}
    \centering
    \includegraphics[width=0.48\textwidth]{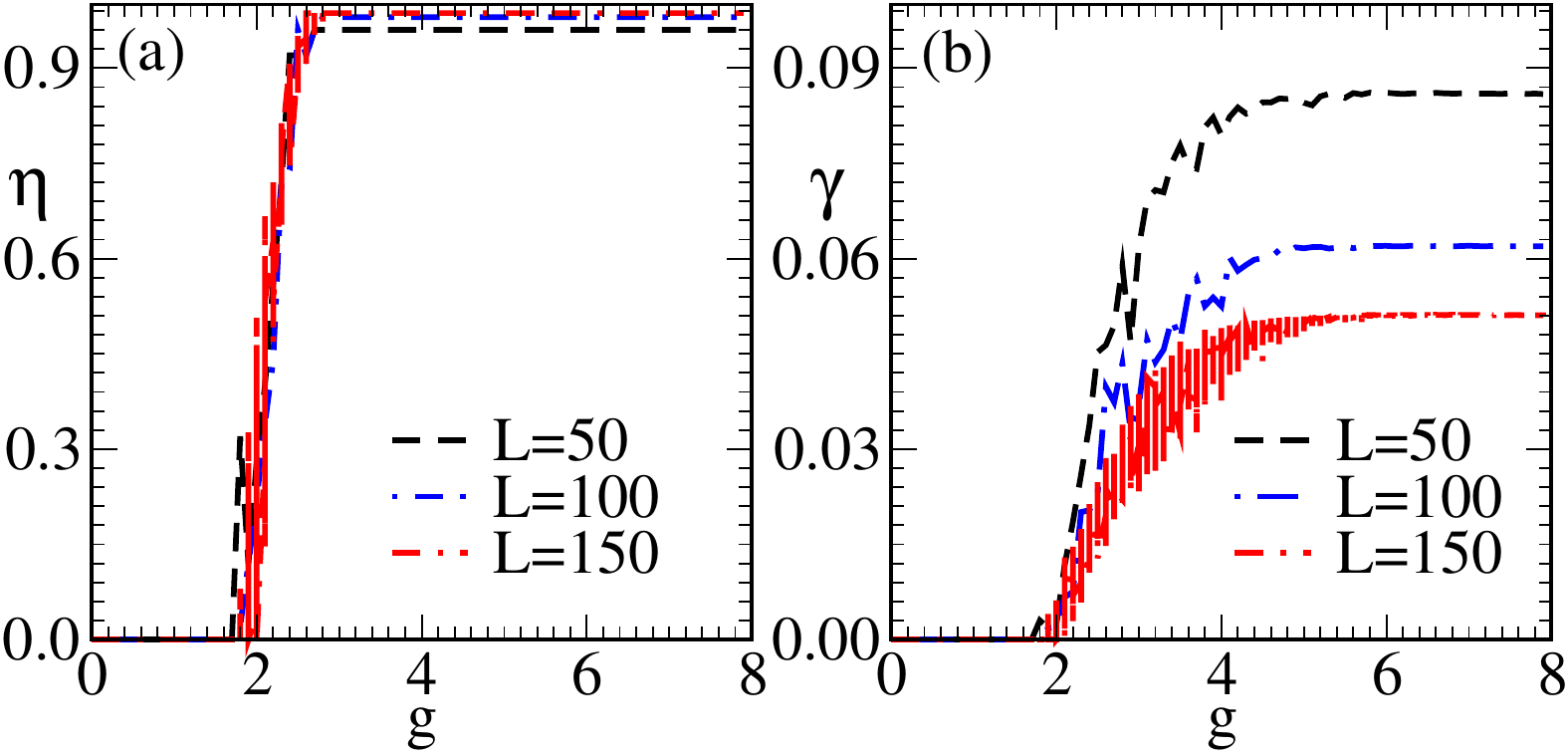}
    \caption{(a) shows the variation of $\eta$ with $g$. (b) shows the variation of $\gamma$ with $g$ for different values of $L=50$, 100, 150, and for Model B. }
    \label{fig2}
\end{figure}
\begin{figure}
    \centering
    \includegraphics[width=0.49\textwidth]{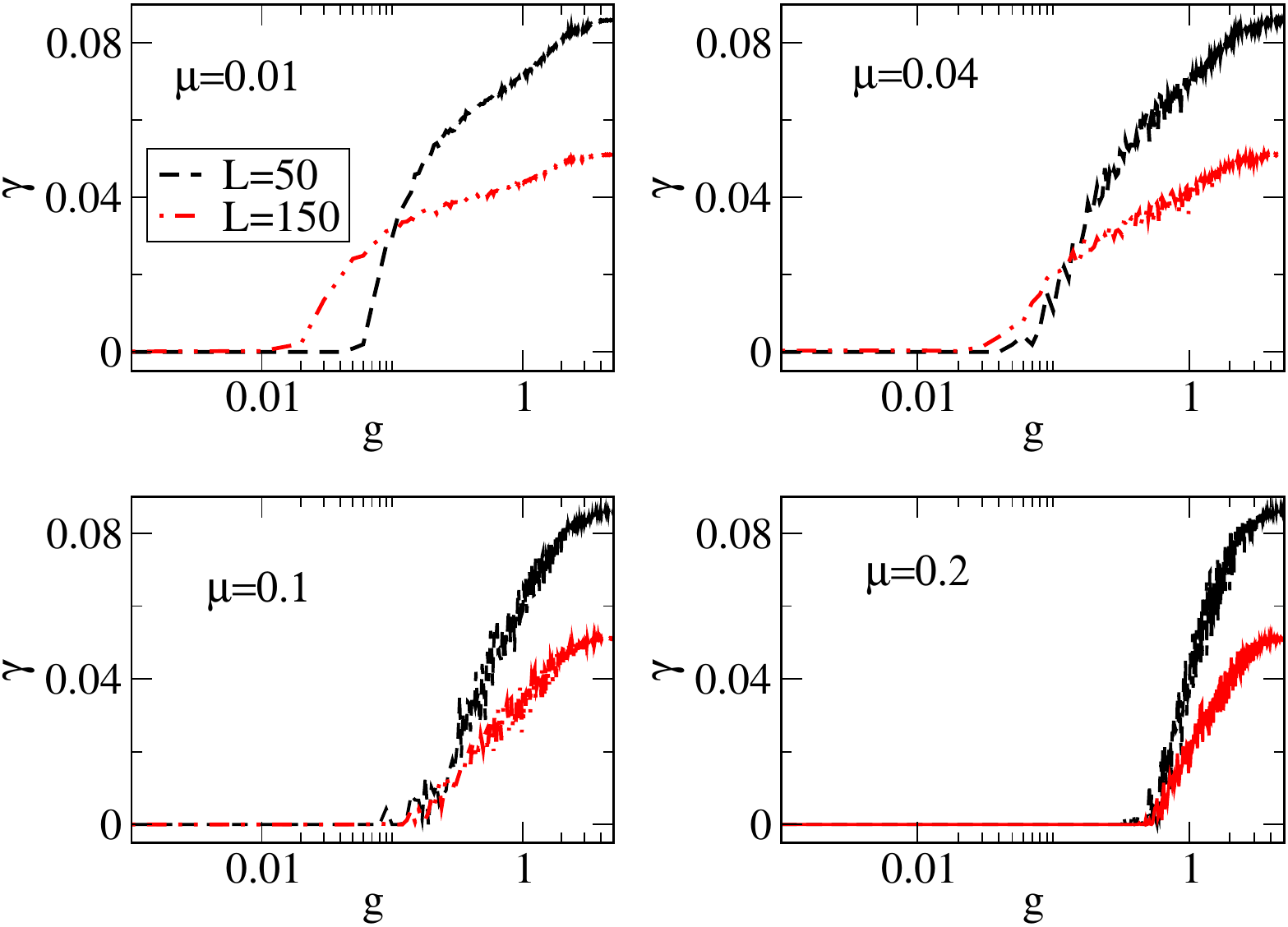}
    \caption{Shows the variation of $\gamma$ with $g$ for different values of $\mu$ and $L$. }
    \label{fig6}
\end{figure}
For Model A, we plot
the fraction of complex eigenvalues $\eta$ as a function of the Hermiticity breaking parameter $g$ in Fig.~\ref{fig1}. For a fixed $L$, there is a critical value of $g_{\rm c}$, below which all eigenvalues are real. This satisfies a necessary condition for the existence of a {\bf PT} transition. To show that the $g$ dependent transition in energy spectrum is indeed a {\bf PT} transition, we need to 
verify if similar transition happens to the corresponding energy eigenfunction properties as well. We define a $\gamma$ measure as,
\begin{eqnarray}
\gamma=\frac{1}{L^2}\sum\limits_{n=1 \atop \langle \text{Re}[R_n]|\text{Re}[R_n]\rangle>0}^{L} \sqrt{\langle \text{Im}[R_n]|\text{Im}[R_n]\rangle} , \nonumber
\end{eqnarray}
where $|R_n\rangle$s are right-eigenvectors of the Hamiltonian $\mathcal{H}_{\rm q}$. Hence, by definition $\gamma=0$ implies that $|R_n\rangle$s are completely real (apart from a trivial global complex phase). 
We find that, coinciding with the transition in the energy spectrum, $\gamma$ is zero for $g<g_{\rm c}$,  and non-zero otherwise, for each $L$ (see Fig.~\ref{fig1}). We thus conclude that model A undergoes a {\bf PT} transition in $g$.  However, $g_c$ is $L$ dependent. We also see from Fig.~\ref{fig1} that as $L\to\infty$, $g_{\rm c}\to0$. This means in the thermodynamic limit thre is no {\bf PT}-transition in model A. Nevertheless, the ground state still shows a spontaneous {\bf PT}- transition even when $L\to\infty$.
In the inset of Fig.~\ref{fig1}, we have studied whether the ground state energy is complex or real by defining a quantity $\eta_{gs}$, which is defined as
$\eta_{gs}=1~\text{when}~\text{Im}[E_{gs}]\neq 0$, and $0$ otherwise. 
We find that there exists a critical value of $g=g_{th}$ (which is $\sim V_0$), above which the ground
state energy is complex, and this $g_{th}$ is independent of $L$. 
However, the value of the imaginary part of the ground state energy, i.e. $\text{Im}[E_{gs}]$ goes to zero as we approach to $L\to \infty$ limit.
In the continuum model, we have also observed the same, and identified it as a DNA unzipping transition as well as a {\bf PT} transition of the ground state. 

For Model B, first, we show the variation of $\eta$ and $\gamma$ with $g$ for $\mu=1$ in Fig.~\ref{fig2}. We see that there exists a $g_{\rm c}$ which separates between the {\bf PT}- symmetric and {\bf PT}- broken phase. Unlike the model A, in this case the {\bf PT}- transition is stable even in the thermodynamic limit because $g_{\rm c}$ does not change as we increase the system size $L$.  However, the essential role is played by the parameter $\mu$ in determining the thermodynamic stability of
the {\bf PT} transition in model B. For $\mu=0$, the model B reduces to the model A.
 We study the parameter regime $0<\mu<1$ in Fig.~\ref{fig6}. By studying the $\gamma$ parameter, we find that the stable {\bf PT} transition in the thermodynamic limit starts occurring for $\mu >0.1$. Also, the critical value $g_c$ increases with $\mu$.

We also investigate localization-delocalization transition, which translates to DNA unzipping in the continuum model, in Model-A and Model-B by using a popular diagnostic tool named `participation ratio'~\cite{zeng.2017,zhou.2022,hui.2019} defined as
\begin{eqnarray}
 \text{PR}_{k}= \frac{\sum_j|\langle L^j_k|R^j_k\rangle|}{\sum_j|\langle L^j_k|R^j_k\rangle|^2},\nonumber
\end{eqnarray}
where $|L^j_k\rangle$ and $|R^j_k\rangle$ are $k$-th left- and right-eigenvector of the non-Hermitian Hamiltonian. We define  
mean participation ratio (MPR) by averaging over all the eigenstates. The localized and delocalized phases of a system 
are indicated by MPR values $\mathcal{O}(1)$ and $\mathcal{O}(L)$, respectively. 
\begin{figure}
    \centering
    \includegraphics[width=0.48\textwidth]{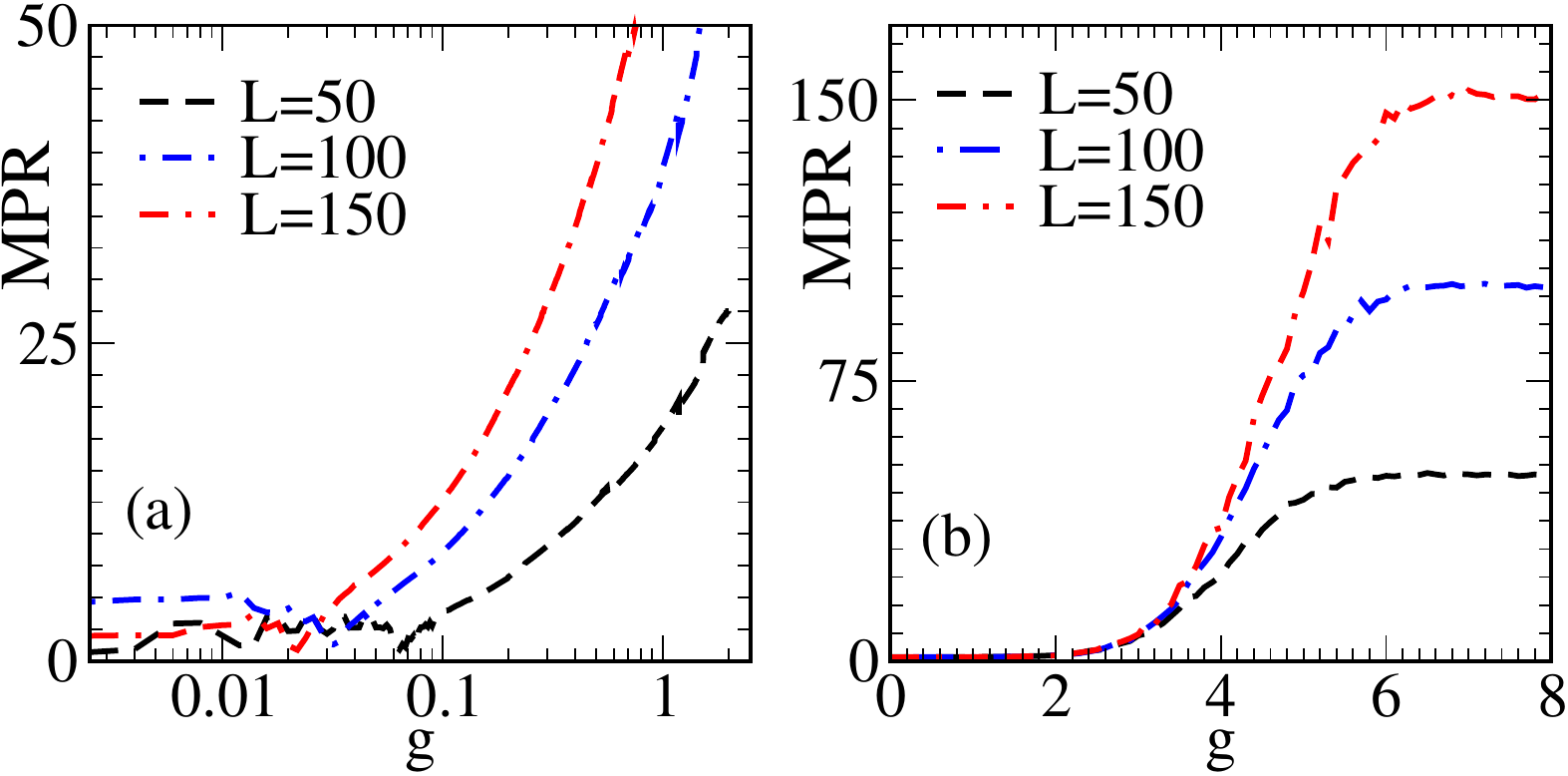}
    \caption{(a) shows the variation of MPR with $g$ for model A. (b) shows the variation of MPR with $g$ for different values of $L=50$, 100, 150 for Model B. }
    \label{fig3}
\end{figure}
Figure.~\ref{fig3} shows the variation of MPR with $g$ for different values of $L$. We see that
MPR remains $\mathcal{O}(1)$ for $g<g_{\rm c}$ and it increases with $L$ for $g>g_{\rm c}$ (ground state $\text{PR}_{k}$ for single impurity model is shown in Appendix-I). MPR values show that 
similar to {\bf PT} transition the localization-delocalization transition persists only for Model-B as $L\to\infty$. The localization-delocalization transition point obtained from the MPR results exactly coincides with the generalized {\bf PT} transition point obtained in Fig.~\ref{fig2}.  This proves that in these models both the unzipping transition and generalized {\bf PT} transition occur simultaneously.

\section{Conclusion} 
This manuscript presents an example of {\bf PT} symmetry breaking transition in an important 
biological system. 
We demonstrate that the one-impurity HN-Hamiltonian, with or without on-site random potential, shows concurring {\bf PT}-breaking and localization-delocalization transitions.
Since localization-delocalization transitions in these models translate to unzipping transitions of DNA and other polymeric systems, we conclude that such unzipping transition is underpinned by spontaneous {\bf PT}-symmetry breaking. In this context, it is interesting to point out that for a single impurity Hatano-Nelson model, in a straightforward way one can show that as one tunes $g$ across its
critical value, $g_c$, the localization length $\xi$ diverges as $\xi\sim|g-g_c|^{-\nu}$
and the excitation gap closes as $|g-g_c|^z$ with the correlation length and dynamic critical exponents $\nu=z=1$. These critical exponents match exactly with the one-dimensional transverse field Ising model, where the ground state shows quantum phase transitions between ferromagnetic and paramagnetic phases 
due to the spontaneously breaking of  a discrete $Z_2$ the spin-flip symmetry
~\cite{PhysRevX.4.031008,dutta2010quantum, sachdev1999quantum}.
Identification of the nature 
of symmetry breaking in case of DNA (or even other polymers) unzipping transition might be helptful to formulate a simple {\bf PT}-symmetric Ginzburg-Landau type 
free energy in future. With the recent advancement in experimental techniques, unzipping force could be applied to a dsDNA single molecule in a very controlled manner using optical tweezers, at ambient conditions \cite{BOCKELMANN20021537,Bocklemann1997PRL, ClaudiaDanilowicz1, ClaudiaDanilowicz2, EssevazRoulet, Anselmetti}. 
The measured end-to-end distance between two DNA strands as a function of force in these single molecule studies 
has the same behaviour as our delocalization measuring observable MPR.  The efforts to directly realize HN-systems in the experimental platforms~\cite{nh_exp1,nh_exp2,nh_exp3,nh_exp4,nh_exp5} turned out to be challenging as it typically requires an external energy source/sink,
and can potentially lead to instabilities due to extreme sensitivity to boundary condition~\cite{instability1,instability2}.
The HN-models with asymmetric kinetic energy term show interesting topological features distinct from Hermitian realm~\cite{nh_exp4}. Topological features originate from bulk-boundary correspondence which in turn translate to 
periodic-open -boundary correspondence. The mathematical equivalence between these quantum HN-systems to 
dsDNA systems would help us designing room temperature experimental set ups e.g. unzipping an dsDNA molecule 
adsorbed on a cylindrical surface. The pulling force in such a system could be applied along the curved surface and the 
radius of the cylinder could be varied compared dsDNA length to get the effects of periodic or open boundary conditions. Moreover, in reality, DNA sequences often have non-uniform distributions of base pairs, and regions with more GC pairs would require significantly higher unzipping forces than AT-rich regions. Such scenarios cannot be captured by the models we have studied here. Hence, looking for more sophisticated, fine-tuned models that can capture these effects is a potential future direction of our work.

\section{Acknowledgements}
RM acknowledges the DST-Inspire research grant by the Department of Science and Technology, Government of India, SERB start-up grant (SRG/2021/002152). 
BPM acknowledges the research grant for faculty under IoE Scheme (Number 6031) of Banaras Hindu University, Varanasi.

\appendix
\section{Ground state participation ratio for the single impurity model}
\begin{figure}[h!]
    \centering
    \includegraphics[width=0.48\textwidth]{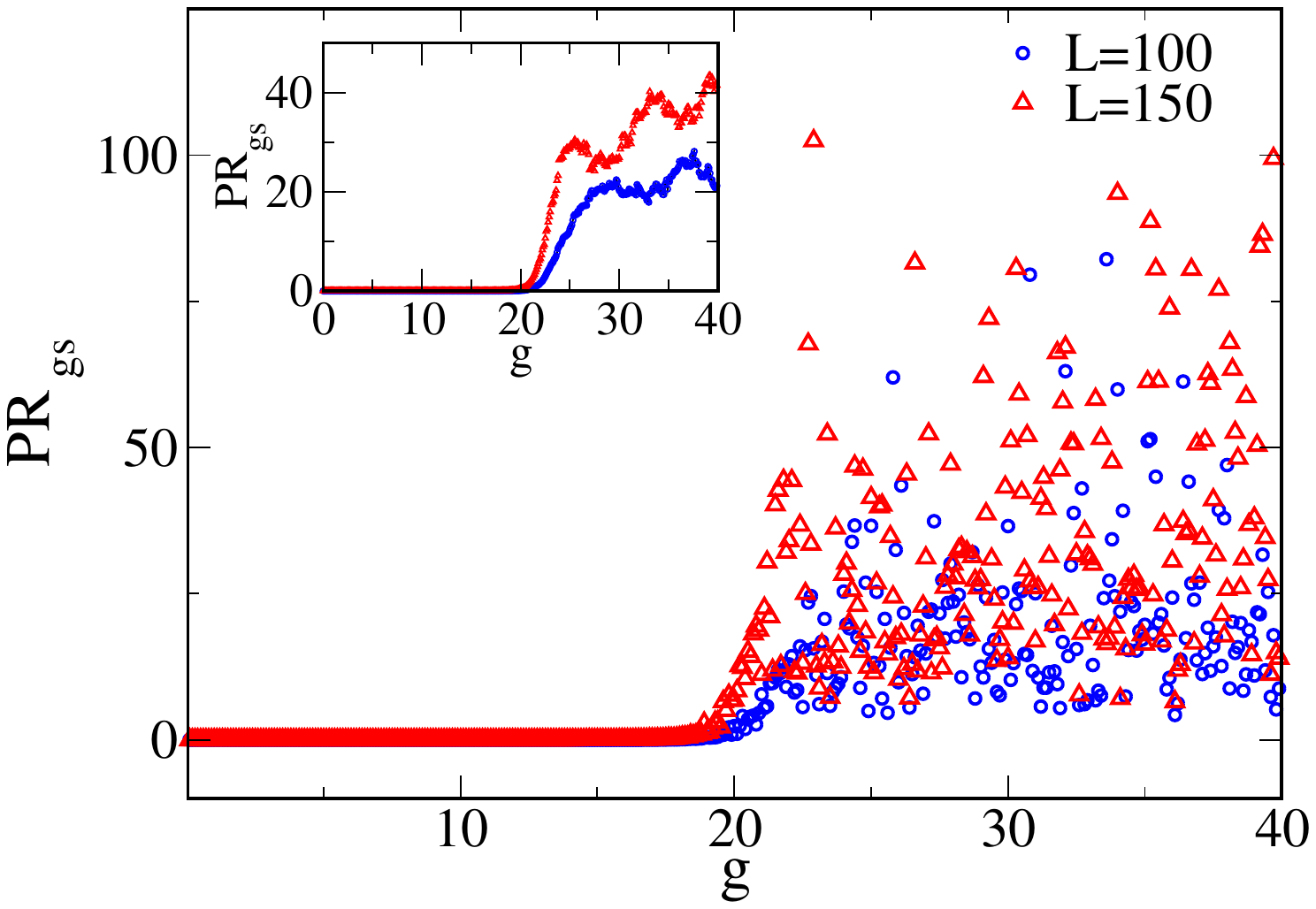}
    \caption{Ground state $\text{PR}_{gs}$ for single impurity discrete model without random potential. Inset shows same results for $\mu=0.0001$, averaged over $1000$ disorder realizations.}
    \label{fig:PR_ground_state}
\end{figure}
Previously, we have shown from the mean participation ratio calculation (where we considered all the states) that for the single impurity model, the transition point $g_c$ approaches zero as we increase the system size $L$. These results are complemented by the real to complex transitions in the eigenvalue spectrum. However, we have found that the ground state energy goes through the real-complex transition at 
$g_c\simeq 20 =V_0$, and this transition point does not change with $L$. In this section, we have also calculated the participation ratio for the ground state and find that it does not scale with $L$ for $g<g_c\simeq 20$, and for $g>g_c\simeq 20$, it increases with $L$. This confirms the existence of localization-delocalization transition in the ground state in the single impurity discrete model at $g_c\simeq 20$.
\bibliography{cite}

\begin{thebibliography}{60}
\bibitem [{Com()}]{Comment1}%
  \BibitemOpen
  \href@noop {} {\bibinfo  {journal} {Signatures of any phase transition can be
  obtained from the free energy per unit chain length at the thermodynamic
  limit which is given by $f= -\lim_{N\to\infty}k_B T N^{-1}\ln Z$}\
  }\BibitemShut {NoStop}%
\end{thebibliography}

\end{document}